\documentclass[preprint,5p,times,twocolumn]{elsarticle}
\usepackage{amssymb}
\usepackage{epsfig,graphics,color}
\usepackage{graphicx}
\usepackage{dcolumn}
\usepackage{bm}
\usepackage{amsmath,eufrak,mathrsfs}

\newcommand \tie {{\it i.e.}}

\newcommand \kd  {\delta}
\newcommand \ra  {\rightarrow}

\newcommand \x {\cdot}

\newcommand \A {\alpha}

\newcommand \lc {\langle}
\newcommand \rc {\rangle}

\newcommand \sg {\sigma}

\newcommand \bvec{\left( \begin{array}{c} }
\newcommand \evec{\end{array} \right)}

\newcommand \bea{\begin{eqnarray} }
\newcommand \eea{\end{eqnarray} } 
\newcommand \nn {\nonumber}
\newcommand {\be} {\begin{equation}}
\newcommand {\ee} {\end{equation}}

\newcommand {\mbx} {\mbox{}}

\newcommand {\ata} {& \times &}

\begin{document}

\begin{frontmatter}



\title{Drag Induced Radiation and Multi-Stage Effects in Heavy-Flavor Energy Loss}


\author[a1]{S.~Cao}
\author[a1]{ A.~Majumder}
\author[a2] {G.-Y.~Qin}
\author[a3]{ C.~Shen}

\address[a1]{Department of Physics and Astronomy, Wayne State University, Detroit, Michigan, USA.}

\address[a2]{Institute of Particle Physics and Key Laboratory of Quark and Lepton Physics (MOE),
Central China Normal University, Wuhan, China.}

\address[a3]{Department of Physics, Brookhaven National Laboratory, Upton, New York, USA.}

\begin{abstract}
It is argued that heavy-quarks traversing a Quark Gluon Plasma, undergo a multi-stage modification process, similar to the case of light flavors. 
Such an approach is applied to heavy-quark energy loss, which includes a rare-scattering, multiple emission formalism at 
momenta large compared to the mass (sensitive only to the transverse diffusion coefficient $\hat{q}$), 
and a single scattering induced emission formalism (Gunion-Bertsch) at momenta 
comparable to the mass of the quark [sensitive to $\hat{q}$, and the longitudinal drag ($\hat{e}$) and diffusion ($\hat{e}_2$) coefficients]. 
The application of such a multi-stage approach within a 2+1D viscous fluid-dynamical 
simulation leads to simultaneous agreement with experimental data for the nuclear modification factor of both $B$ and $D$ mesons, as measured by the CMS 
collaboration at the LHC. The extracted transport coefficients are found to be consistent with those for light flavors.
\end{abstract}


\end{frontmatter}


\section{Introduction}
\label{intro}

The logarithmic dependence of the QCD coupling constant $[\A_S (Q^2)]$ on the scale of the process ($Q^2$) ensures that phenomena that depend on strong interactions change dramatically as the scale changes by an order of magnitude~\cite{Gross:1973id,Politzer:1973fx}. In the collision of heavy-ions at the Relativistic Heavy-Ion Collider (RHIC) and the Large Hadron Collider (LHC), the Quark Gluon Plasma (QGP) is expected to have a maximal temperature of $T_{Max}$$\sim$$0.5$~GeV~\cite{Song:2011qa}, which is only slightly higher than the vacuum scale $\Lambda_{QCD} \sim 0.2$~GeV. 
In contrast, light flavor jets at the LHC span a range from $50$~GeV up to $500$~GeV~\cite{Aad:2012vca}, with leading hadrons measured up to $300$~GeV\cite{Khachatryan:2016odn}. 

Heavy flavor suppression, measured as the nuclear modification factor of $D$ and $B$ mesons, has now been measured by the Compact Muon Solenoid (CMS) collaboration up to $50$~GeV~\cite{Sirunyan:2017oug,Sirunyan:2017xss}~(the experimental data in these papers forms the genesis of the current Letter).
Prior to these measurements, there have been a variety of experimental estimates of heavy-flavor loss, from non-photonic electron suppression, to suppressed yields of hard 
$J/\psi$ mesons. Given the dead-cone effect for heavy-flavor~\cite{Dokshitzer:2001zm}, particularly when the momenta is not parametrically larger than the mass, one expects less suppression for heavy-flavors, with a mass ordering in the nuclear modification factor.
In all cases, a theoretical explanation of the suppression, as well as the azimuthal anisotropy, has been somewhat of a challenge, especially using the same transport coefficients as extracted from the known suppression of light flavors~\cite{Qin:2009gw}. In this Letter, we will demonstrate that a multi-stage energy loss formalism
along with new mass dependent additive components to radiative loss that depend on the drag and longitudinal diffusion coefficients ($\hat{e}$, $\hat{e}_2$), are sufficient to explain the additional suppression seen in the heavy-flavor sector.

 The $c$ and $b$ quarks that eventually fragment into $D$ and $B$ mesons, start as highly virtual partons from the hard interaction that produces them. In the process of traversing the medium they will lose this virtuality via 
multiple vacuum like and medium induced emission\cite{Majumder:2009zu}. 
Prior to exiting the medium they will undergo a stage of only medium induced emission. Such a multi-stage 
approach was recently applied to the case of light quarks and gluons in Ref.~\cite{Cao:2017zih}. In that effort, the existence of a high virtuality and low virtuality 
stage in the radiation from a hard parton, allowed for a natural mechanism for the medium modified jet shape. 

In this Letter, we devise the first multi-stage formalism for heavy-flavor energy loss: This includes a high energy, high virtuality stage, where, compared to the energy and virtuality, the quark mass can be neglected. In this stage, the heavy-quark undergoes a medium modified~\cite{Wang:2001ifa,Guo:2000nz,Majumder:2009ge} Dokshitzer-Gribov-Lipatov-Altarelli-Parisi (DGLAP) shower~\cite{Dokshitzer:1977sg,Gribov:1972ri,Gribov:1972rt,Altarelli:1977zs}. This depletes the virtuality and energy of 
the heavy-quark. For cases where the resulting remnant momentum of the heavy-quark is comparable to the mass of the heavy-quark, it undergoes further energy loss via a Gunion-Bertsch~\cite{Gunion:1981qs} type of process, where scatterings off the medium trigger radiation from the heavy-quark. Such a stage is to be distinguished from the traditional picture of multiple scattering induced radiation for massless flavors, which requires several scatterings per emission, and is denoted as the Baier-Dokshitzer-Mueller-Peigne-Schiff (BDMPS) approach~\cite{Baier:2001yt,Baier:1998yf,Baier:1996sk,Baier:1996kr,Baier:1994bd}. In this Letter, we will explore the phenomenological implications of the assertion of Refs.~\cite{Abir:2014sxa,Abir:2015hta}, that for semi-hard heavy-quarks  with a
momentum comparable to their mass, there are two important differences with light flavor energy loss: (i) There is a mass dependent contribution of drag and longitudinal diffusion to radiative energy loss; (ii) semi-hard heavy-quarks do not have a BDMPS stage, \tie, the formation time of the radiation is parametrically shorter than for massless flavors, leading to the aforementioned Gunion-Bertsch stage of single scattering per emission.
 
 In this Letter, we will demonstrate that the above two effects lead to a different ``radiative'' energy loss for heavy-quarks compared to light flavors; this is above and beyond the extra presence of drag and diffusion terms in the elastic energy loss of heavy-quarks. In Sec.~\ref{semihard}, we will review the theoretical description of radiative loss from semi-hard heavy-quarks. In Sec.~\ref{raa} we will describe the matching of the two different approaches to heavy-flavor loss, in effect, how one mechanism of energy loss transitions to another. This is followed by an outline of the phenomenological aspects of the calculation of the nuclear modification factor, and comparison with new experimental data from CMS. Concluding discussions are presented in Sec.~\ref{disc}.

\section{Semi-hard heavy-quarks}
\label{semihard}

It is now almost universally accepted that drag and diffusion play an important role in the energy loss of heavy-quarks in a QGP. 
The drag is often referred to as elastic energy loss, though, it is only elastic as far as the projectile is concerned, \tie, the scattering of a 
hard quark leads to a reduction of its forward momentum and energy, but may include an in-elastic disintegration of a degree of freedom 
in the target~\cite{Majumder:2008zg}. However, until very recently, the possibility that longitudinal scattering, which leads to drag and longitudinal diffusion, 
can change the off-shellness of the hard parton, resulting in a change to the radiation rate, has not been explored. 

The calculations in the current Letter, are based on two sets of papers which both explored the effect of drag and diffusion 
coefficients on stimulated emission from a hard massless and massive quark: In Refs.~\cite{Qin:2012fua,Qin:2014mya,Zhang:2016avg} the authors considered the effect of the light-cone drag and longitudinal diffusion coefficients, $\hat{e} = d\Delta p^-/dL^-$ and $\hat{e}_2 = d (\Delta p^-)^2/dL^-$~\cite{Majumder:2008zg}, on radiative loss from a massless quark. They 
observed the effect of $\hat{e}$ and $\hat{e}_{2}$ on radiative loss to be parametrically suppressed, compared to the effect of $\hat{q}$.  
Contemporarily, the authors of Refs.~\cite{Abir:2014sxa,Abir:2015hta} studied the effect of these transport coefficients on the radiative loss from a massive quark. For the case of massive quarks, in particular for quarks where the momentum $p$ is comparable to the mass of the quark $M$, both $\hat{e}$ and $\hat{e}_2$ produce mass dependent 
corrections to the gluon emission cross section. For $M$ large enough, these corrections could become comparable to the effect of $\hat{q}$ on the single gluon emission cross section. 

\subsection{The case of massless flavors}

We start with a familiar process of a hard and light quark or gluon, produced in a hard scattering, beginning to propagate through a QGP medium, say in the $-z$-direction, towards a target propagating with a comparable boost factor, in the $+z$ direction. To get an estimate of scales, we are considering the energy to be of the order of $E\sim100$~GeV. This undergoes transverse scattering off several gluons in the medium, with each gluon possessing a 
transverse momentum of $k_\perp \sim 1$~GeV. To distinguish this large difference in scales, we use the scaling variable $\lambda$, and the standard notation of
$Q$ denoting the hard scale. However, our method of using $\lambda$ will differ from the usual setup in Soft-Collinear-Effective-Theory (SCET)~\cite{Bauer:2000yr,Bauer:2002nz,Bauer:2001yt,Bauer:2001ct}. We pick the momentum of the hard parton to scale as $q \equiv (q^+, q^-, \vec{q}_\perp)\sim Q(\lambda,1,\sqrt{\lambda})$. We make this choice to make direct contact with the power counting scheme in Ref.~\cite{Abir:2015hta}. 
The main advantage of this choice is that it makes it manifest that $\sqrt{\lambda}$ terms are not ignored compared to $\lambda^0$ terms. 
We will eventually only retain terms that scale as $\lambda$ or greater. In the phenomenological situation that we are considering, $Q \sim 100$~GeV, and $\lambda \sim 0.01$. 
One could repeat our analysis with the choice of $\sqrt{\lambda}$ replaced as $\lambda$ and then only retain terms which scale as $\lambda^2$ or greater, obtaining the same results. 
The gluon from the medium scales as $k \sim (\lambda^2, \lambda^2, \lambda) Q$. These are called 
Glauber gluons~\cite{Idilbi:2008vm}. Within this scaling, the difference between the scales in a hard vacuum scattering and the softer scales in the medium is 
manifest.

Following the scaling rules above, the off-shellness of the hard parton is $q^2 = \lambda Q^2$. This leads to a formation time of the radiated 
gluon of,
\be
\tau = \frac{2E}{q^2} \sim \frac{1}{\lambda Q}.
\ee
(One could also define a light-cone formation time $\tau^- = 2q^-/q^2$, which is of the same order.)
At the location of each scattering, one evaluates the thermal expectation~\cite{Majumder:2012sh}:
\be
\mbx\!\!\!\!\!\int  \!\!\!d y^-  \!\!\! \int  \!\!d \kd y^-  e^{i k^+ \kd y^- + i \kd k^+ y^- }\left\langle F^{+ \perp} \left( y^- + \frac{\kd y^-}{2} \right)  F^+_ \perp \left( y^- - \frac{\kd y^-}{2} \right) \right\rc,
\ee
where, $k^+$ and $\kd k^+$ represent the mean and uncertainty in the (+)-component of the exchanged momentum, and $y^-$ and $\kd y^-$ represent the 
mean and uncertainty of the location of the scattering. The brackets $\lc \rc$ denote a local thermal expectation in an extended QGP.
For the assumption of independent scatterings to hold true $y^- \gg \kd y^-$ and $k^+ \gg \kd k^+$.  One should note that $k^+$ and $y^-$ are not conjugates of each other and thus 
can be simultaneously large compared to $\kd k^+$ and $\kd y^-$. 
Given the range of $k^+$ which will not change the off-shellness of the hard parton, coupled with the requirement that the mean distance between scatterings ($y^-$) has to be larger than the uncertainty, we 
would obtain that $y^- \sim 1/\lambda Q$. Thus the mean number of scatterings encountered by the hard parton within the formation time are $N = \tau^- /y^- \sim 1$.  This is the regime of medium modified DGLAP, where there is at most one scattering per emission.

As has been demonstrated in Ref.~\cite{Majumder:2014gda}, after traveling some distance in a dense medium, the hard light parton has lost a considerable portion of its virtuality and some of its momentum and energy, and now possesses a momentum $q \sim (\lambda^2, 1 , \lambda )Q$.  One notes in both cases that the
transverse momentum is closely related to the virtuality of the parton. The formation time of 
radiation is now, 
\be
\tau = \frac{1}{\lambda^2 Q}.
\ee
Over this time, the number of scatterings encountered can be as large as $N=\tau^-/y^- \sim 1/\lambda$. 
This is the regime of BDMPS radiation, where radiations are separated by longer separations, induced by multiple scattering per emission.
As such, for the case of light flavors, one notes two distinct stages: One which we indicate as the medium-modified DGLAP stage, with multiple emissions with 
the occasional scattering in between. This is followed by the BDMPS stage with fewer well separated emissions with multiple scattering per radiation.

\subsection{The case of massive flavors}

We now consider the case of massive flavors, 
where the mass of the heavy-quark is intermediate between the hard scale $Q$ and the soft scale $\lambda Q$, \tie, $M \sim \sqrt{\lambda} Q$.  
For the case of high momentum massive quarks, formed in a hard interaction, the four momentum components scale as $q\sim (\lambda,1,\sqrt{\lambda})Q$.
At these scales the quark behaves similar to the case of a massless flavor. We are interested in the case of semi-hard heavy-quarks, where the momentum is 
comparable to the mass of the quark, \tie, $q^- \sim \sqrt{\lambda} Q$. For $B$ and $D$ mesons in the $10$-$50$~GeV range, this is of particular relevance to $b$-quarks. 

Consider the case of a semi-hard heavy-quark which has propagated some distance in the medium, and passed through the medium modified DGLAP stage. 
These now have momentum components $q \sim (\sqrt{\lambda} , \sqrt{\lambda} , \lambda) Q$.  Since the heavy-quark's largest momentum component scales as $\sqrt{\lambda} Q$, it cannot radiate a gluon with momentum of order $Q$. The analysis of Ref.~\cite{Abir:2015hta} suggests that the highest scaling 
of the radiated gluon is $l \sim (\lambda, \lambda, \lambda) Q$.  The transverse momentum has to remain of of order $\lambda Q$ as it is generated by the scattering off the Glauber gluon, which also carries a transverse momentum of order $\lambda Q$. The scaling of $l^+$ is driven by the requirement that the 
radiated gluon be on-shell. 
The formation length of such radiation, is 
\be
\tau_Q = \frac{2 l^-}{ l^2} \sim \frac{1}{\lambda Q},
\ee
comparable to the short formation times of the DGLAP regime (one should not estimate this from the momentum components of the heavy-quark as those include the mass). In this time, the number of scatterings is of order unity. Thus heavy-quarks even in the stage where the large virtuality at origin has been radiated away, undergo one scattering per emission. It is this scattering that drives the heavy-quark off shell and causes it to radiate. We have referred to this regime as the Gunion-Bertsch regime, due to the similarity of the process with that in Ref.~\cite{Gunion:1981qs}.

The real part of medium modified portion of the single gluon emission cross section,  in this stage, depends not only on the transverse diffusion coefficient $\hat{q}$ but also on the longitudinal drag and diffusion coefficients 
$\hat{e}$ and $\hat{e}_2$. The approximate form of this cross section, expanded up to terms suppressed by $\lambda$, can be expressed as, 
\bea
\mbx\!\!\!\!\!\!\!\!\!\!\frac{dN}{dydl_\perp^2 d \zeta^-} \!\!\!&=&\!\!\! \frac{2 C_F \A_S}{2\pi} 
\frac{P(y)}{\left( l_\perp^2 + y^2 M^2 \right)^2} 2 \sin^2\left( \frac{l_\perp^2 + y^2 M^2}{ 4 l^- y(1-y)} \zeta^-\right) \label{NewFormula} \\
\ata\!\!\! \left[ \hat{q} \left\{  1 - \frac{y}{2} - \frac{y^2 M^2}{l_\perp^2}\right\} + \hat{e} \frac{y^2 M^2}{l^-} + \hat{e}_2 \frac{y^2 M^2}{2(l^-)^2}\right]. \nn
\eea
In the equation above, $y$ is the light cone momentum fraction radiated away by the gluon $y=l^-/q^-$ where $q^-$ is the large light-cone momentum fraction 
of the heavy-quark which has a mass of $M$. The factors $C_F$ and $\A_S$ are the Casimir and strong coupling constant. The splitting function $P(y)$ is the 
regular Altarelli Parisi splitting function~\cite{Altarelli:1977zs}, and $\zeta^-$ is the light-cone location of the scattering. One immediately notes the new mass dependent 
terms that enhance the effect of $\hat{e}$ and $\hat{e}_2$ on the single gluon emission rate, only for the case of massive quarks. As $M\ra 0$, these terms diminish, yielding 
the expression for the single gluon emission rate for light quarks. 

In the subsequent section, we will use the above equation to estimate the energy loss for semi-hard heavy-quarks. For hard virtual heavy-quarks, we will use the standard 
expressions for the medium modified DGLAP equation used for light flavors.

\section{Nuclear modification factor for D and B mesons}
\label{raa}

In this Letter, we present the first realistic calculation of the suppression of $B$ and $D$ mesons using a two-stage process where high momentum quarks 
$p_Q \gg M$, undergo medium modified DGLAP evolution, similar to light quarks, followed by a stage of lower momentum quarks undergoing a Gunion-Bertsch like energy loss. Calculations are carried out within the framework of factorized pQCD, where we can express the cross section of heavy-hadrons in bins of $p_T$ and 
rapidity $y$, within a range of centrality (between $b_{min}$ to $b_{max}$), as 
\bea
\mbx\!\!\!\frac{d \sigma^{AA}(p_T,y)_{b_{min}}^{b_{max}} }{dy d^2 p_T} 
\!\!\! &=& K \int_{\mbx_{b_{min}}}^{\mbx^{b_{max}}}\!\!\!\!\!\!\!\!\!d^2 b 
\int d^2 r  T_{AB}(\vec{r}, \vec{b})  \nn \\
\!\!\! &\times& \!\!\!\! \int  d x_a d x_b  G^A_a(x_a,Q^2)  G^B_b(x_b,Q^2)  \nn \\
\ata \!\!\!\! \frac{  d \hat{\sg}_{ab \ra cd} }{ d \hat{t}}  
\frac{ \tilde{D}_Q^h(z,Q^2, \hat{p}_{Q} )}{\pi z}.  \label{AA_sigma}
\eea
In the equation above, $T_{AB}$ represents the thickness function for the two colliding nuclei. The nuclear parton distribution functions of the two colliding 
nucleons to produce partons $a$ and $b$, \tie, $G^A_a(x_a,Q^2)$ and $G^A_b(x_b,Q^2)$ are evaluated at the hard scale $Q=p_T$ of the detected hadron.
The hard partonic cross section $\hat{\sg}$ is evaluated at the partonic Mandelstam variable $\hat{t}$. The entire two stage energy loss of the 
hard parton leading to the fragmentation into a $D$ or $B$ meson is contained within the medium modified fragmentation function $\tilde{D}_Q^h$ evaluated 
at the momentum fraction $z$, evolved up to a scale of $Q=p_T$, with an original parton momentum of $\hat{p}_Q$. 

The calculation of the medium modified fragmentation function starts with the vacuum fragmentation function, measured using the PYTHIA event generator.  
The medium modification is calculated in two parts: A modification of the vacuum fragmentation 
function using Eq.~\eqref{NewFormula}, followed by an in-medium DGLAP evolution.
In this work, heavy-quarks are initialized with a MC-Glauber model for their spatial distribution and a leading order pQCD calculation for their momentum space distribution. The CTEQ6 parametrization~\cite{Pumplin:2002vw} is adopted for the parton distribution functions  for their momentum space distribution and the EPS09 parameterization~\cite{Eskola:2009uj}  is used to estimate the amount of nuclear shadowing effect in the initial state. 

Evolution of heavy-quarks below a separation scale $Q_0$, inside the QGP, is described using a time-ordered transport model -- an improved Langevin approach~\cite{Cao:2011et,Cao:2013ita,Cao:2015hia}, in which the elastic scattering process is calculated with the standard Langevin equation for the Brownian motion. Inelastic scattering processes follow a rate equation derived from Eq.~\eqref{NewFormula}. The rate of gluon emission at the location $\zeta^-$ is obtained as, 
\begin{equation}
\label{eq:gluonnumber}
\Gamma_\mathrm{rad}(\zeta^-) = \int dy \int dl_\perp^2 \frac{dN}{dy dl_\perp^2 d\zeta^-}.
\end{equation} 
Here, the transverse and longitudinal diffusion coefficients in Eq.~\eqref{NewFormula} are related by dimensionality $\hat{q}=2\hat{e}_2$. The longitudinal drag and diffusion coefficients are related by the Einstein relation $\hat{e}=\hat{e}_2/(2T)$. The QGP medium is simulated with a (2+1)-D viscous hydrodynamic model~\cite{Song:2007fn,Song:2007ux,Qiu:2011hf,Shen:2014vra}, that is initialized with the MC-Glauber model and provides the space-time evolution of the local temperature and fluid velocity of the QGP fireballs. With this information, for every time step, we boost each heavy-quark into the local rest frame of the fluid cell through which it propagates and update its energy and momentum based on our transport approach. The heavy-quark is then boosted back to the global center of mass frame where it streams freely until its interaction with the medium in the next time step. After heavy-quarks travel outside the QGP hypersurface, they are converted into heavy flavor mesons through PYTHIA simulations, and their medium modified fragmentation function is extracted. Note that we constrain this transport approach to apply only at a low $Q^2$ scale by requiring both the $l^2_\perp$ in Eq.~\eqref{eq:gluonnumber} and the parton virtuality in the PYTHIA shower to be smaller than $Q^2_0$. The lower limit of the $y$ integration is derived from the ratio of the thermal scale to the momentum of the quark; the upper limit depends on the allowed kinematics for a given energy of the heavy-quark. 

\begin{figure}[h!]
$\mbx$
\hspace{0.25cm}
\includegraphics[width=0.4\textwidth]{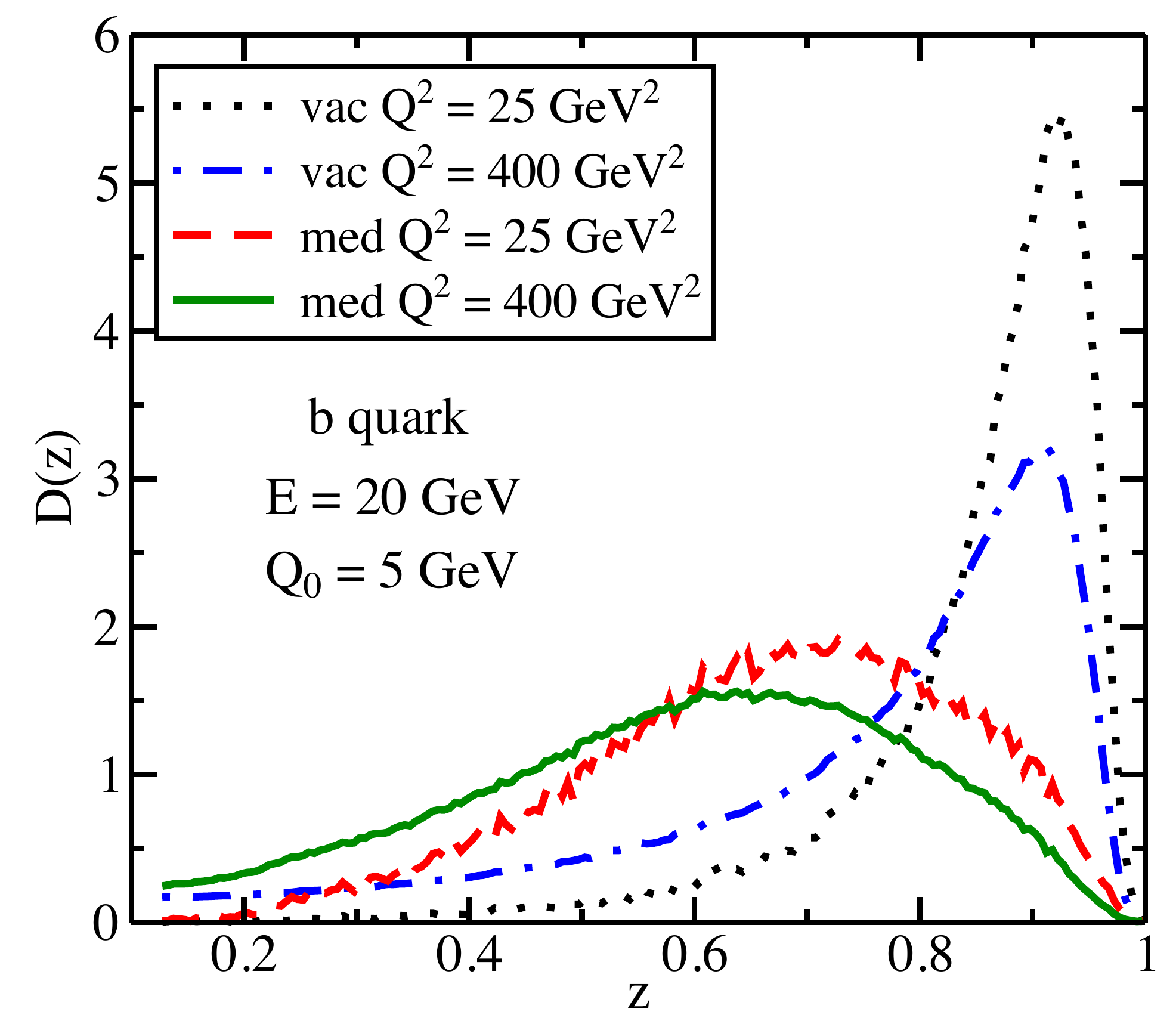} 
\vspace{-0.5cm}
    \caption{The b to B fragmentation function as measured at $Q_0=5$~GeV in vacuum and then evolved via Eq.~\eqref{NewFormula}, to a medium modified fragmentation function at the scale $Q_0=5$~GeV, then further 
    modified via the medium modified DGLAP equation to the hard scale $Q=20$~GeV. This is compared with vacuum evolution from from $Q_0$ to $Q$.}
    \label{BmesonFragFuncEvolution}
\end{figure}

The medium modified fragmentation functions at the intermediate scale $Q^2_0$ have both $z$ and energy dependence, as the evolving equation [Eq.~\eqref{NewFormula}] contains an energy dependence, in addition to the $z$ dependence in the vacuum Altarelli-Parisi (AP) equation. Different evolution equations are used to evolve the fragmentation functions from 
$Q_0^2$ to the hard scale $Q^2 = p_T^2$. In vacuum this is carried out using the standard AP equations, with the momentum fraction $z$ defined using the light-cone momentum, \tie, with mass included in the definition. In the case of evolution in the medium we use the higher twist medium modified DGLAP evolution equation~\cite{Wang:2001ifa}. It should be pointed out that for $p_T^2 \sim Q_0^2$ the effect of the medium modified DGLAP equation is minimal compared to the evolution below the scale $Q_0$. 
The medium modified DGLAP equation will have a noticeable effect when $p_T \gg Q_0$. In the calculations presented in this Letter, we have consistently set $Q_0 \sim M_{\rm meson}$ \tie, comparable to the mass of the detected meson ($2$~GeV for $D$, $5$~GeV for $B$).

\begin{figure}[h!]
$\mbx$
\hspace{.2cm}
\includegraphics[width=0.45\textwidth]{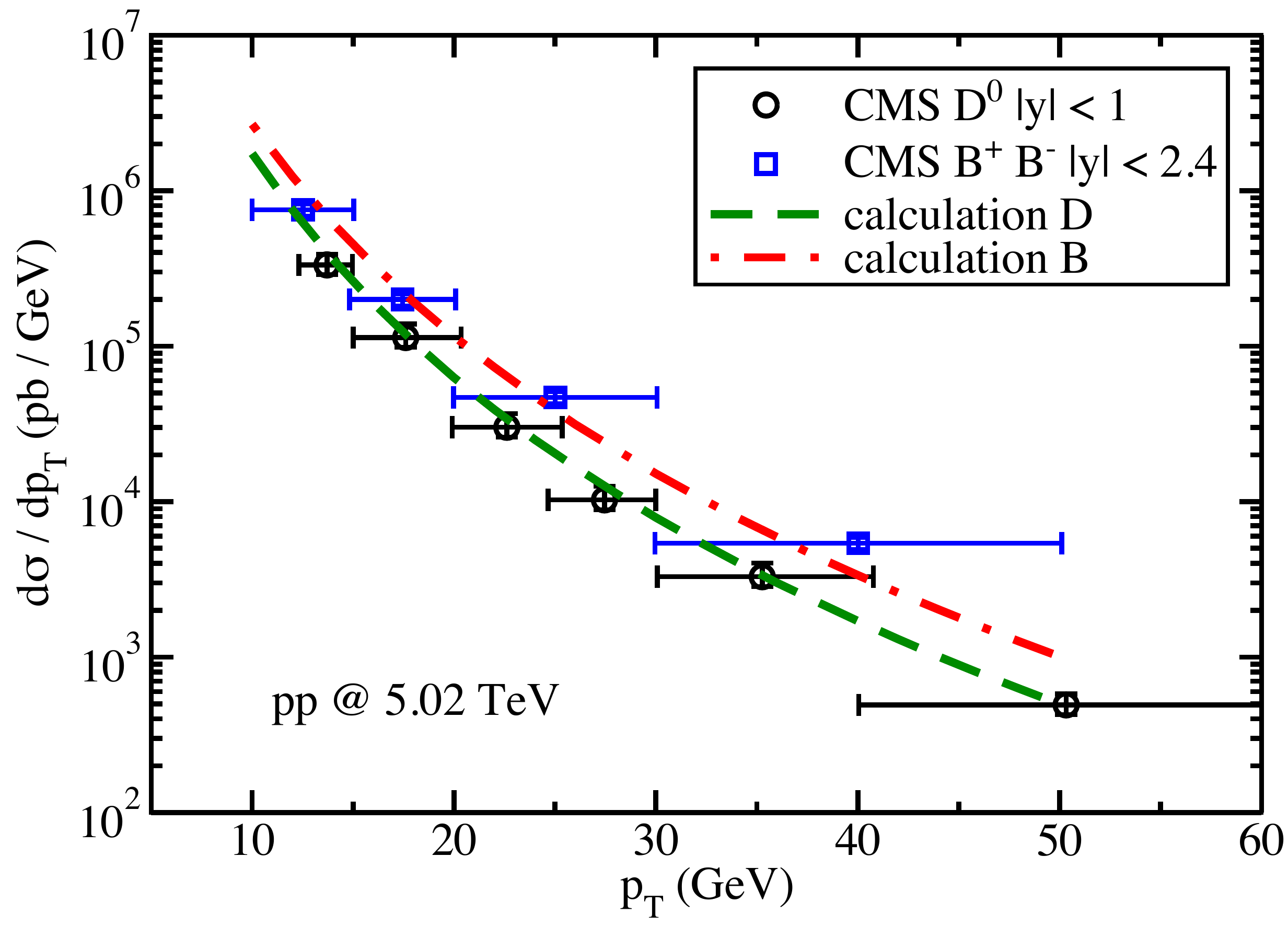} 
\vspace{-0.5cm}
    \caption{The spectrum of $B$ and $D$ mesons adjusted with $K$ factor to account for NLO corrections, and multiplicative 
    factors to account for the multitude of mesons detected. }
    \label{BandDmesonSpectraPP}
\end{figure}

The effect of the different evolution equations on the fragmentation functions are included in Fig.~\ref{BmesonFragFuncEvolution} for a $B$ meson fragmenting from a $b$ quark . The vacuum fragmentation functions are measured using the PYTHIA event generator, by isolating a $b$ quark with energy $E=20$~GeV, in a $p$-$p$ collision and then evolving up to $Q=p_T$. 
The spectrum obtained by convoluting this fragmentation function with our parametrized hard scattering cross section and parton distribution functions is presented in Fig.~\ref{BandDmesonSpectraPP}. These calculations involve a $K$-factor of $1.7$ in the $p_T$ range explored.
As can be seen, we obtain a good description of the vacuum spectrum of $B$ and $D$ mesons in $p$-$p$ collisions. 

The in-medium evolution, as presented in Fig.~\ref{BmesonFragFuncEvolution} for a $b$-quark, involves a production point distributed according to the binary collision profile of a minimum bias $Pb$-$Pb$ collision. The direction of propagation is distributed randomly in $\phi$ at $\eta=0$ (we are using a $2+1$~D viscous fluid dynamical simulation). The vacuum fragmentation function for this quark undergoes evolution via the transport equation to obtain medium modified function at $Q_0$. 
It is then evolved up to the hard scale using a medium modified DGLAP evolution equation.

\begin{figure}[h!]
$\mbx$
\hspace{.2cm}
\includegraphics[width=0.45\textwidth]{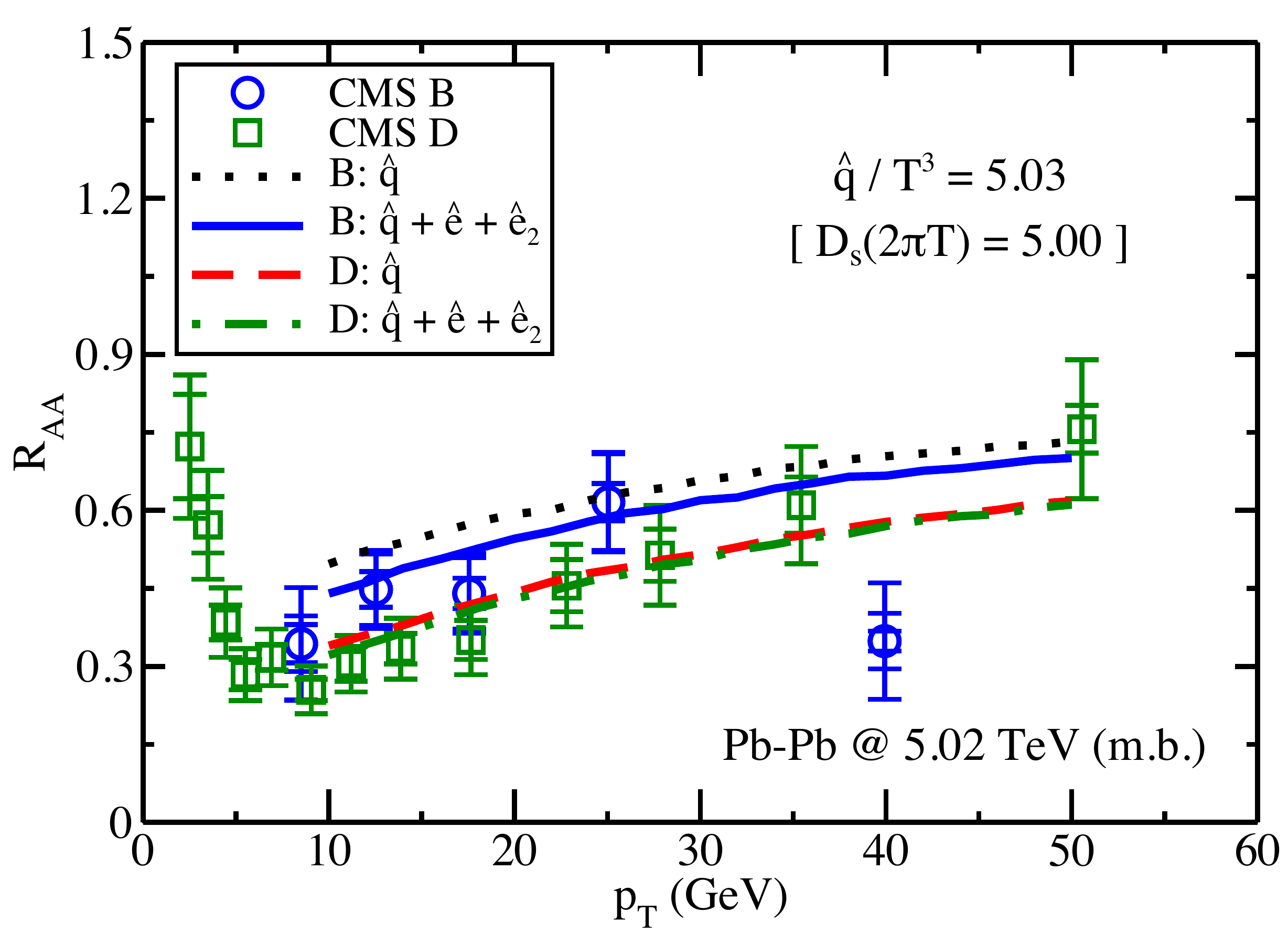} 
\vspace{-0.5cm}
    \caption{The nuclear modification factor for $D$ and $B$ mesons as measured by the CMS collaboration. The lines are theoretical 
    calculations based on the inclusion of gluon radiation induced by transverse diffusion, both with and without the effect of longitudinal drag and diffusion on the radiated gluon. All lines include transverse diffusion, longitudinal drag and diffusion for the non-radiative portion of the evolution. The medium modified DGLAP evolution from $Q_0$ to $Q$ only includes the effect of transverse diffusion.}
    \label{BandDmesonRaa}
\end{figure}

In Fig.~\ref{BandDmesonRaa} we calculate the nuclear modification factor for $D$ and $B$ mesons and compare to minimum bias data, as measured by the $CMS$ collaboration. 
The transport coefficients are dependent on the local temperature $T$ and are related to each other.
The value of $\hat{q}$ is obtained from the normalization equation $\hat{q} = 5.03T^3$. 
One will note that this value is within one standard deviation from the central value at LHC energies, obtained for light flavors by the JET collaboration~\cite{Burke:2013yra}. 
We obtain simultaneous agreement with the experimental data for both $D$ and $B$ meson 
nuclear modification factors, using a $\hat{q}$ which is similar to that used for light flavors. 
In order to test the importance of the mass dependent terms in Eq.~\eqref{NewFormula}, 
we also present calculations where the effect of $\hat{e}$ and $\hat{e}_2$ are removed from Eq.~\eqref{NewFormula}. 
As would be expected these have a noticeable effect on the $B$ meson $R_{AA}$ and much less on the 
$D$ meson $R_{AA}$. 
This is because, at a $p_T \gtrsim 10$~GeV, the $c$ quark is effectively a light quark. 
One will also note that these mass dependent effects diminish for larger energies. 
It should be expected that at higher energies, when $p\gg M$, the heavy-quarks will also engender a BDMPS like stage at lower virtualities, 
and a medium modified DGLAP stage at higher virtualities, similar to light flavors.

\section{Discussions and conclusions}
\label{disc}

In conclusion, we have presented the first implementation of a multi-stage formalism to heavy-quark energy loss. 
The suppression observed for heavy-flavors in comparison with light flavors, in light of the dead-cone effect, has remained 
somewhat at odds with theoretical expectations: There has always seemed to be more suppression than expected, 
given that the mass of the quark would curtail collinear emission, which plays a central role in the calculated larger energy 
loss of light flavors. In this Letter, we continue the proposal of Refs.~\cite{Abir:2014sxa,Abir:2015hta}, that the process of radiative energy 
loss of heavy-quarks is quantitatively different from light flavors for two related reasons. Similar to the case of light flavors, 
hard heavy-quarks start from a hard scattering far off mass shell and lose virtuality via a medium modified DGLAP equation.
However, unlike the case of massless flavors, semi-hard heavy-quarks exist in a state where their momentum is larger but comparable 
to the mass of the heavy-quark. 
In this state, heavy-quarks lose energy by a combination of drag and radiative loss via a Gunion-Bertsch like process of single scattering induced emission. 
The single gluon emission rate for semi-hard heavy-quarks, depends not only on the transverse momentum diffusion coefficient $\hat{q}$, but also on the 
longitudinal drag and diffusion coefficients $\hat{e}$ and $\hat{e}_2$, through mass dependent terms that vanish for massless quarks. 

In this Letter, we have made the first realistic estimate of the nuclear modification factor for $B$ and $D$ mesons at the intermediate momentum range from 10-50~GeV. The expectation is that the $b$ and $c$ quarks that fragment into these, do so in the vacuum after escape from the medium. We measure the vacuum 
fragmentation functions using the PYTHIA event generator, and then evolve these in a 2+1~D viscous hydrodynamic medium. The vacuum fragmentation functions are measured at the minimal 
scale of $Q_0 = 2$~GeV for $c \ra D$, and $Q_0 = 5$~GeV for $b \ra B$ fragmentation. They are evolved in the medium, first using Eq.~\eqref{NewFormula} from the 
starting location of the hard scattering. They start as semi-hard quarks and undergo a Langevin like energy loss process where they lose energy by drag, diffusion and radiative loss.
They are then evolved form $Q_0$ up to the hard scale $Q$ using a medium modified DGLAP equation. These medium modified fragmentation functions at the hard scale 
$Q =p_T$ are then convoluted with a parametrized hard scattering cross section to obtain the final spectum of $B$ and $D$ mesons and the nuclear modification factor. 
Using a $\hat{q}/T^3$ which is within one standard deviation from the results of light flavors as obtained from the JET collaboration, we obtain good agreement with the 
experimental data for $B$ and $D$ nuclear modification factors, in minimum bias events at $5.02$~TeV. 
This Letter offers justification that both heavy and light flavors can be described using the same overarching framework based on pQCD, with the difference that there are 
mass dependent contributions to heavy-quark energy loss, which are vanishing for the case of light flavors. These extra contributions lead to an enhancement of the radiative 
loss rate, compensating for the reduction due to the dead-cone effect. 
Future improvements on this formalism will require calculations to be done within a multi-stage Monte-Carlo event generator as developed by 
the JETSCAPE collaboration~\cite{Cao:2017zih}, where the medium modified DGLAP stage, the Gunion-Bertsch and BDMPS  phases can be applied over each
path travelled by the hard quark. 

\section{Acknowledgements}
The authors thank R.~Abir for helpful discussions regarding his earlier work. The work of S. C. and A. M. is supported in part by the U.S. Department of Energy (DOE) under grant number 
\rm{DE-SC0013460} and in part by the National Science Foundation (NSF) within the framework of the JETSCAPE collaboration, under grant numbers \rm{ACI-1550300}.
G.-Y. Q. is supported by the NSFC of China, under grant numbers 11375072 and 11775095.
C.S. gratefully acknowledges a Goldhaber Distinguished Fellowship from Brookhaven Science Associates, his work is also supported in part by the U.S. Department of Energy (DOE) under grant number \rm{DE-SC0012704}.


\bibliographystyle{elsarticle-num} 
\bibliography{refs}





\end{document}